\documentclass[journal]{IEEEtran}

\ifCLASSINFOpdf
\else
\fi

\usepackage[cmex10]{amsmath}
\usepackage{amssymb}
\usepackage{hyperref}
\usepackage{gensymb}
\usepackage{textcomp}
\usepackage{pifont}
\usepackage{graphics}
\usepackage{cite}

\usepackage{float}
\usepackage{amsmath}
\usepackage{amstext}
\usepackage{amsfonts}
\usepackage[pdftex]{graphicx}
\usepackage{makecell}
\usepackage{array,multirow}
\usepackage{color}

\usepackage{mathtools}
\usepackage{booktabs,tabularx}

\newcolumntype{C}[1]{>{\centering\arraybackslash}p{#1}}

\usepackage{flushend}

\begin{document}
%
\title{Domain Generalization for Document Authentication against Practical Recapturing Attacks}
%
%
%

\author{Changsheng Chen,~\IEEEmembership{Senior Member,~IEEE},
        Shuzheng Zhang,~\IEEEmembership{Student~Member,~IEEE},
        \\Fengbo Lan,~\IEEEmembership{Student Member,~IEEE},
        Jiwu Huang,~\IEEEmembership{Fellow, IEEE}

\thanks{The authors are with the Guangdong Key Laboratory of Intelligent Information Processing and Shenzhen Key Laboratory of Media Security, and National Engineering Laboratory for Big Data System Computing Technology, College of Electronics and Information Engineering, Shenzhen University, Shenzhen, China. They are also with Shenzhen Institute of Artificial Intelligence and Robotics for Society, China (e-mail: cschen@szu.edu.cn, jwhuang@szu.edu.cn).}
}

\maketitle

\begin{abstract}
Recapturing attack can be employed as a simple but effective anti-forensic tool for digital document images.
Inspired by the document inspection process that compares a questioned document against a reference sample, we proposed a document recapture detection scheme by employing Siamese network to compare and extract distinct features in a recapture document image.
The proposed algorithm takes advantages of both metric learning and image forensic techniques.
Instead of adopting Euclidean distance-based loss function, we integrate the forensic similarity function with a triplet loss and a normalized softmax loss.
After training with the proposed triplet selection strategy, the resulting feature embedding clusters the genuine samples near the reference while pushes the recaptured samples apart.
In the experiment, we consider practical domain generalization problems, such as the variations in printing/imaging devices, substrates, recapturing channels, and document types.
To evaluate the robustness of different approaches, we benchmark some popular off-the-shelf machine learning-based approaches, a state-of-the-art document image detection scheme, and the proposed schemes with different network backbones under various experimental protocols.
Experimental results show that the proposed schemes with different network backbones have consistently outperformed the state-of-the-art approaches under different experimental settings.
Specifically, under the most challenging scenario in our experiment, i.e., evaluation across different types of documents that produced by different devices, we have achieved less than 5.00\% APCER (Attack Presentation Classification Error Rate) and 5.56\% BPCER (Bona Fide Presentation Classification Error Rate) by the proposed network with ResNeXt101 backbone at 5\% BPCER decision threshold.
\end{abstract}

\begin{IEEEkeywords}
Document Image, Recapture Detection, Deep Learning,
\end{IEEEkeywords}

\section{Introduction}
\label{sec:Intro}

\begin{figure}
\centering{
\begin{minipage}[c]{.425\linewidth}
  \centering
  \centerline{\includegraphics[height=0.95in]{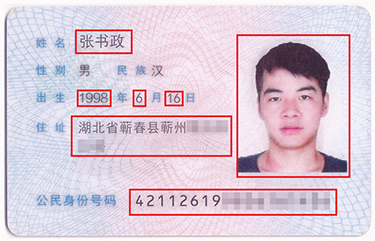}}
  \centerline{\footnotesize(a)}
\end{minipage}
\hspace{0.25cm}
\begin{minipage}[c]{.425\linewidth}
  \centering
  \centerline{\includegraphics[height=1.15in]{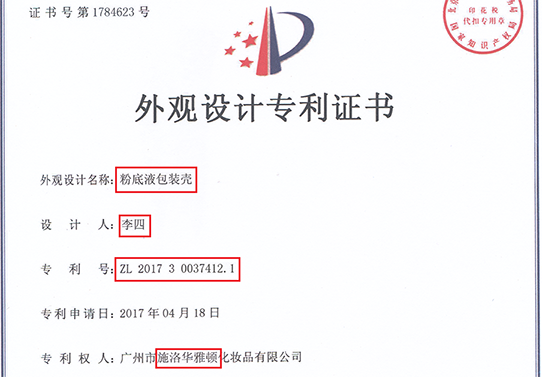}}
  \centerline{\footnotesize(b)}
\end{minipage}
}
\caption{Examples of recapturing attack on document images. The forgery trace is concealed by the recapturing (reprint and re-acquiring) operation. (a) A tampered and recaptured identity document (ID) image. Sensitive information has been processed by mosaic. (b) A tampered and recaptured patent certificate document image (an enlarged view). The printing and scanning devices in generating (b) are Canon G3800 and Canoscan 5600F, respectively.}
\vspace{-0.35cm}
\label{fig:IDCardSpoof}
\end{figure}

There is urgent demand of online document authentication with increasing online services in the e-commerce and e-government applications.
Authentication of digitally acquired documents image from hardcopy documents is an important forensic task with broad interest.
However, the security loophole in the exiting authentication scheme has put our system at risk.
As shown in Fig.~\ref{fig:IDCardSpoof}, the text in an identity document (ID) image and a certificate image can be edited by some off-the-shelf image editing softwares (e.g., Adobe Photoshop) or some latest deep learning-based techniques \cite{zhao2021deep}.
To cover the forgery trace, the edited ID image can be re-acquired through a print-and-scan cycle.
The resulting images in Fig.~\ref{fig:IDCardSpoof} are therefore more realistic than those edited in digital domain.
It should be noted that the deep learning-based editing techniques proposed in \cite{zhao2021deep} requires few human intervention, and the recapturing attack can be carried out by non-expert attackers using commodity printers and scanners.
Thus, the document forgery attack together with recapturing operation has posed a new threat to our document authentication system.

Detection of document recapturing attack is the key step of identifying document from the forgery-and-recapture attack. However, recapturing detection of document image in practical applications is difficult since there are several variations in the recapturing process.
First, the attacker may employ different imaging devices (such as a mobile/point-and-shot/professional camera), instead of a dedicated scanner.
According to our preliminary study, a low-quality mobile imaging device introduce different capturing distortions compared to a high-quality scanner.
Second, printing process with different printing techniques (such as inkjet, laserjet) with distinct halftoning patterns (such as screening, dithering) and substrates (such as office paper, high-quality glossy paper) should be considered.
Third, the recapturing attack may be launched by completely different procedure.
For instance, instead of the recapturing attack through the print-and-scan channel, the attacker may display the forged document on an LCD screen and re-acquire it with a camera, i.e., attack through the display-and-capture channel.
Last but not least, there is a large variation of the document types in a practical document authentication system. It includes identity cards, certificates, contracts, etc.
In summary, there is significant domain shift between samples collected under different settings.
The main difficulty of document recapturing detection resides in generalizing the performance towards samples in different domains.

There are various research works on document image authentication.
The existing techniques can be divided into the active and the passive categories.
As an example of the active forensic techniques, the digital watermarking techniques \cite{cox2007digital} can be applied to a document as protection against illegal alternations or re-acquisition.
However, the active techniques demand controls over the document generation process, which limits its application to documents from various parties.
In contrary, the passive techniques authenticate a document by the intrinsic features without the need to embed additional signals.
For instance, a questioned document image can be examined through some inherent characteristics of the printing and acquisition process \cite{chiang2009printer, mayer2020forensic} for tampering detection.
However, the existing passive forgery detection techniques on digital images are not robust under the recapturing attack.
Under such an attack, the tampered image of a given document is output with a printed and re-acquired with an imaging device to generate a recaptured version of the image.
The recaptured document image has been through a complete image acquisition chain, and no post-processing (or forgery) after the acquisition steps.
This image will be considered as an original copy by the existing passive tampering detection techniques.
To fight against such attacks, the image recapturing detection has attracted global research attentions.
However, most of the exiting recapture detection schemes focus on natural images and biometric images, only a few preliminary works have considered the recapturing attack on hardcopy documents \cite{berenguel2019recurrent, shang2014detecting, Alibaba2020}.
Worse still, as demonstrated in our experiment, the existing document recapturing detection schemes are not robust under domain variations in practical scenarios, such as printing/imaging devices, printing substrates, recapturing channels, and document types.

This paper focuses on the domain generalization problem for document authentication under recapturing attacks.
Instead of tackling this problem in a general image authentication setting, we consider document recapturing detection in a verification setup with the aid of a genuine document image in the same template.
Though requiring additional information in the authentication process, we want to emphasize two important points.
First, such document authentication process that compares a questioned document with a reference is more intuitive according to the forensic process by a human expert.
Second, the proposed scheme is practical for authenticating important documents (e.g., ID cards and patent certificates) with publicly available templates.
With the above idea in mind, the proposed method takes advantage of the recent advancements in metric learning by constructing distinctive triplets (containing genuine, recaptured and reference document images) with considerations on the document contents and acquisition devices.
The features of samples in a triplet are extracted by generic CNN backbones with shared weights.
The distances between features in a triplet are then measured by an end-to-end trainable forensic similarity network, instead of employing the traditional Euclidean metric.
The proposed loss integrates the forensic similarity function with a triplet loss and a normalized softmax loss.
By optimizing the overall loss, a feature embedding space generated by the proposed scheme clusters the genuine samples while pushes the recaptured samples apart.

To evaluate the effectiveness of our approach, a database consists of 132 captured and 972 recaptured document images are collected.
We consider a wide variety of experiment settings, including different printing/imaging devices, different printing substrates, different recapturing channels, and different types of documents.
Experimental results show that the proposed schemes with different CNN backbones have consistently outperformed other approaches under various experimental settings.
Specifically, under the most challenging scenario in our experiment, i.e., evaluation across different types of documents that produced by different devices, the proposed network has achieved less than 5.00\% APCER (Attack Presentation Classification Error Rate) and 5.56\% BPCER (Bona Fide Presentation Classification Error Rate) by the proposed network with ResNeXt101 backbone at 5\% BPCER decision threshold.

The main contributions of this work can be summarized as follows.
\begin{itemize}
\item We propose a recaptured document authentication scheme by verification against a reference image using the Siamese network architecture. Instead of adopting the general Euclidean distance, the proposed network employs a trainable forensic similarity subnet to measure the distance in the embedding space.
\item We take advantage of the recent advancements in metric learning and construct triplets with considerations on the document image contents, acquisition conditions, and resolutions. Distinctive features embedding space can therefore be learnt from triplets with various templates and different resolutions.
\item We construct a practical document images database covering authentication scenarios with different printing/imaging devices, printing substrates, recapturing channels, and document types. The proposed scheme consistently demonstrates strong domain generalization performances under different experiment protocols.
\end{itemize}

The remaining of this paper is organized as follows.
Section~\ref{sec:Literature} reviews the related literatures on document recapturing detection and existing document databases.
Section~\ref{sec:Proposed} introduces the proposed network architecture, forensic loss in the feature embedding space, triplet sampling strategy, and visualizes the feature embedding space.
Section~\ref{subsec:Database} elaborates the data collection procedures of the captured and recaptured document images.
Section~\ref{subsec:Result} compares the performance of different approaches under different settings, including intra-dataset, cross-dataset and in-the-wild experiments.
Section~\ref{sec:Conclusion} concludes this paper.

\section{Literature Review}
\label{sec:Literature}

\subsection{Related Works on Document Recapturing Detection}
\label{subsec:LiteratureRecapture}

Existing image recapture detection schemes mainly focus on three types of media, i.e., biometric images, natural images, and document images.
However, the biometric spoofing detection schemes aim at detecting the liveness of the inspected image.
For example, in the problem of face spoofing detection, researchers extract the forensic trace from rPPG signals (Remote photoplethysmography, i.e., heart pulse signal) and depth \cite{sun2020face}.
It should be noted that these features are not applicable to document images which are printed on flat paper surface.

The research of recapture detection algorithms focus on natural images starts as early as 2010 \cite{cao2010identification}.
It shows that the original and recaptured images (from LCD screens) are not easily distinguishable by naked eye.
Thongkamwitoon \emph{et al.} detected the recaptured images based on edge blurriness and distortion which can be characterized by K-SVD dictionaries from a single image.
Mahdian \emph{et al.} presented a spoofing detection scheme with the periodic properties present in the LCD recaptured images.
Li \emph{et al.} proposed a hierarchical data-driven approach for image recapturing detection by extracting the intra-block information by CNN and capturing the inter-block dependency by recurrent neural network (RNN).
Yang \emph{et al.} proposed a Laplacian CNN model for small-size recapture image forensics \cite{yang2016recapture} and a set of quality-aware feature to detect images recaptured from LCD screens \cite{yang2017recaptured}.
Anjum and Islam extracted the trace of image recapturing by exploiting the high-level details present in images, such as edge profile \cite{anjum2019recapture}.
However, the above works focus on identifying recaptured natural images from LCD screens, which is different from spoofing detection of hardcopy document images.
Both the captured and recaptured versions of a hardcopy document is acquired from flat paper surfaces, which lacks the distinct differences between a 3D natural scene versus a flat and pixelated LCD screen.
Worse still, to control production cost, a generic document is usually produced without employing special printing substrates and technologies.

There are only a few literatures on detecting spoofing document images.
Shang \emph{et al.} built a set of discriminate features from the characteristics of laser/inkjet printers and electrostatic copiers \cite{shang2014detecting}.
The features were then employed to distinguish documents produced by printing and copying devices.
Over 90\% classification accuracy has observed in a dataset collected with 10 laser printers, 6 inkjet printers and 9 copiers.
Berenguel \emph{et al.} adapted a recurrent comparator architecture with attention mechanism to detect forged banknote and ID document with recapturing operation \cite{berenguel2019recurrent}.
It iteratively inspects different textual region of the document to distinguish a counterfeit document with an authentic one.
Xu in Alibaba \cite{Alibaba2020} proposed a recaptured detection scheme with the aid of built-in flash of the mobile phones. The document images are assumed to be acquired under a predetermined lighting condition such that the uncertainties in the capturing environment can be reduced.
However, these works have not considered the domain variation in practical applications, such as different printing/imaging devices, printing substrates, recapturing channels, and document types.

\subsection{Existing Document Databases}
\label{subsec:LiteratureDatabase}

A publicly available database is important resource for benchmarking the performance of different approaches, but there is only a few document image datasets collected for the problem of recapturing detection.
On the one hand, most recapture image datasets focus on natural images \cite{cao2010identification, agarwal2018diverse} and biometric images \cite{sun2020face}.
On the other hand, most document image datasets were built for other applications, e.g., optical character recognition \cite{arlazarov2019midv, bulatov2020midv}, image classification \cite{harley2015evaluation}, source device identification \cite{rabah2020supatlantique}, and document forgery detection in digital domain (without recapturing) \cite{sidere2017dataset}.

Recently, Shruti \emph{et al.} established a database of 14,500 rebroadcast images captured from hundreds of different devices through crowdsourcing \cite{agarwal2018diverse}.
The images are acquired by scanning a printed photo, rephotographing a displayed/printed photo.
However, due to the uncontrolled nature of the sample collecting environment, the samples in this database are not reflecting the scenarios where an attacker tries to fool the document authentication system with careful operations.
The quality of samples in this dataset vary significantly even in the same type of attack.
Moreover, the content of the images mainly includes human subjects and natural scenes, which are very different from that processed by a document authentication system.

A dataset of 6615 banknote and 7550 legal ID document images from different countries is established in \cite{bulatov2020midv}.
The dataset contains both original image and counterfeit document images produced by the scanning-printing operation.
However, this dataset is not publicly available due to the sensitive content. Similar non-publicly available datasets have also been reported in \cite{berenguel2019recurrent, berenguel2017evaluation}.

Given the literatures presented in Section~\ref{subsec:LiteratureRecapture} and \ref{subsec:LiteratureDatabase}, it can be concluded that there are important limitations of the existing works.
Firstly, the generality of existing document image recapture detection algorithms \cite{shang2014detecting, berenguel2019recurrent, Alibaba2020} towards various settings of the recapturing attack is remained to be investigated. According to our experimental results in Section~\ref{subsec:Result}, the state-of-the-art approach \cite{berenguel2019recurrent} is not performing well under some practical settings.
Secondly, a publicly available dataset to evaluate different approaches under some representative settings is missing.
Thus, in the following section, we propose a document authentication scheme with potential to identify recaptured document image under practical domain variations and establish a representative dataset (covering scenarios of different printing/imaging devices, printing substrates, recapturing channels, and document types) to benchmark the performance of different approaches.

\begin{figure*}
\centering{
\includegraphics[width=6.5in]{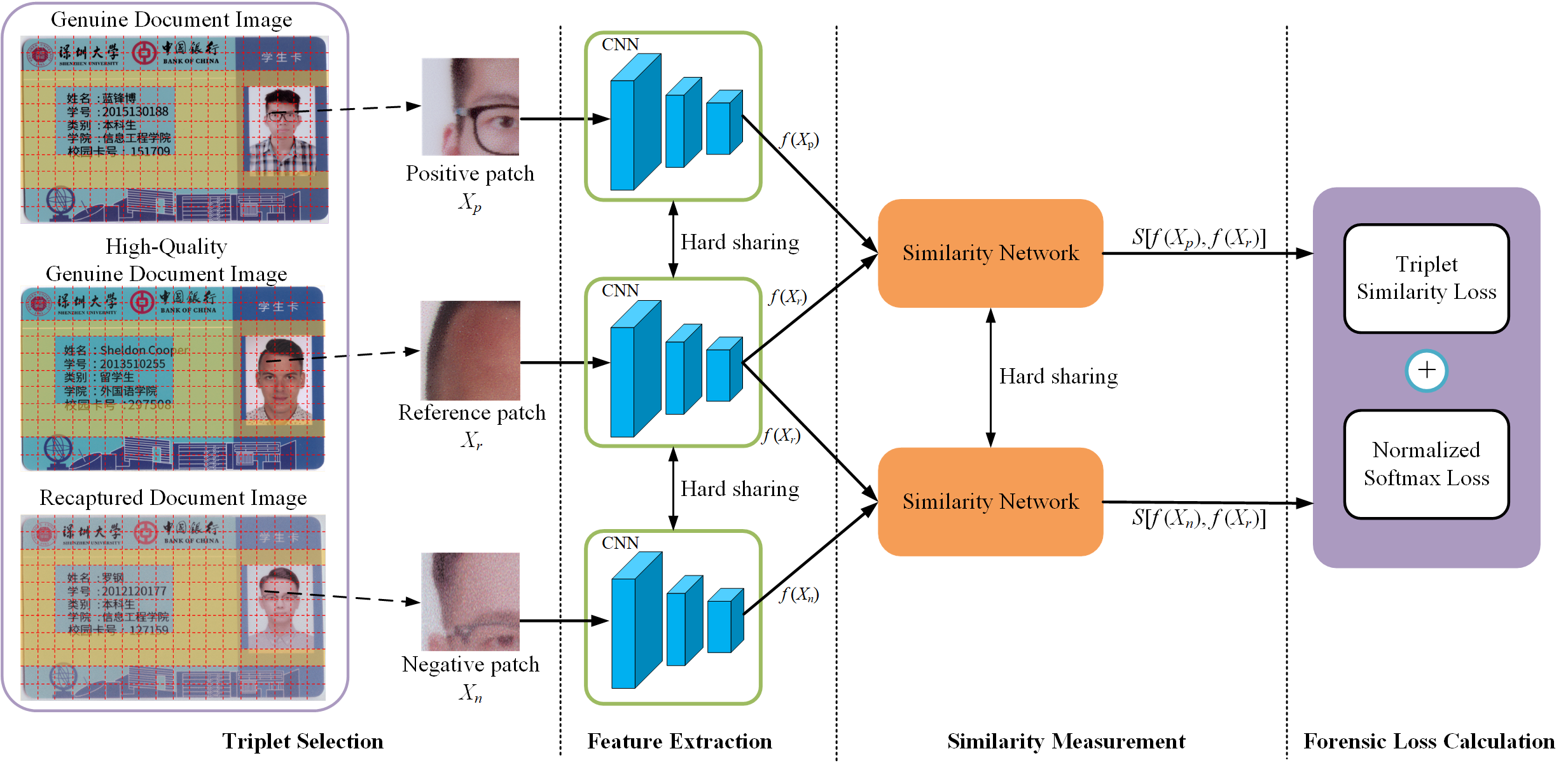}}
\vspace{-0.1cm}
\caption{Network architecture of the proposed recaptured document authentication scheme. The whole process can be divided into four stage, including triplet selection, feature extraction, similarity measurement, and forensic loss calculation.}
\label{fig:NetworkArchitecture}
\vspace{-0.25cm}
\end{figure*}

\section{the Proposed Document Authentication Scheme}
\label{sec:Proposed}

In this work, we consider the document image authentication problem in a different perspective.
Inspired by the fact that a human expert determines the authenticity of a questioned document usually by inspection side-by-side with a genuine document in the same template (or by checking some common features from a template), we believe that it is more intuitive to train a deep network to authenticate a document with the aid of a reference image, such as a high-quality authentic document image in the same template.

Thus, this work addresses the document authentication problem from the verification perspective.
To allow comparing the questioned and reference document images through the network, we adopt a Siamese network architecture which takes two or more images as parallel inputs.
The features of the input images are extracted by the network and compared in the latent domain.
The authentication decision will be determined based on the distance between questioned and reference document images.

Given our analogy to the document inspection procedure of a human expert, vital parts of the proposed authentication scheme reside in extracting distinctive features, comparing features in the latent space and organizing the paired inputs.
Thus, we will elaborate our network architecture for feature extraction, forensics loss in the feature embedding space, and sample paring strategy in the following subsections.

\subsection{The Proposed Network Architecture}
\label{subsec:NetworkStructure}

In this part, we propose a deep learning architecture based on Siamese CNNs and metric learning.
This is to allow multiple samples input to the network at the same time for mimicking the document inspection behavior of a human expert.
The genuine, recaptured, and reference (high-quality genuine) document images in the same template are taken as input.
As shown in Fig.~\ref{fig:NetworkArchitecture}, three patches (extracted from the positive, negative, and reference images, respectively), named a triplet, is input to the network simultaneously.
The positive, negative, and reference patches are denoted as $x_{p_i}, x_{n_i}$ and $x_{r_i}$, respectively, where $i$ is the index of triplet in a mini-batch.
The recapturing traces, such as halftone and color degradation of the input samples are extracted by a set of generic CNNs with shared network weights.
The extracted features are therefore denoted as {\small$f(x_{p_i}), f(x_{n_i})$} and {\small$f(x_{r_i})$}, respectively.
The reference features are paired, respectively, with the features from positive and negative samples, yields two tuples {\small$[f(x_{p_i}), f(x_{r_i})]$} and {\small$[f(x_{n_i}), f(x_{r_i})]$}.
Each tuple is then evaluated by a trainable similarity network to compute distance between the two elements.
By designing a customized loss function, the proposed deep model is trained to yield an embedding space that
minimizes the distance between {\small$[f(x_{p_i}), f(x_{r_i})]$} while maximizes the distance computed from {\small$[f(x_{n_i}), f(x_{r_i})]$}.
Thus, a feature embedding space can be generated to cluster the positive samples near the reference, while pushes the negative samples apart.
The proposed forensic loss in the feature embedding space is discussed in Section~\ref{subsec:ForensicLoss}.
Moreover, randomly choosing positive and negative samples from tens of thousands of possible combinations to form the triplets is neither efficient nor effective.
The details of our triplet sampling strategy are elaborated in Section~\ref{subsec:TripleSampling}.

In the testing procedure, we considers two scenarios depending on whether the template of the questioned document is included in the training data or not.
First, if the template has been seen in the training data, the questioned document image is authenticated by comparing against both the positive and negative samples (of the same template) in the source domain.
The first scenario shows the authentication process of some important documents, such as identity cards and birth certificate, which is in an official and publicly available template.
Second, if the template is new to the training data, several pairs of positive and negative samples in the target domain would be required in the authentication process.
This can be considered as a few shot learning scenario where only a few samples in the target domain are available.
For both cases, the authenticity of a questioned document can be determined by the relative distances among the questioned, positive and negative samples.
The distance metrics from three triplets (including three pairs of positive and negative samples, as well as a questioned sample) are evaluated to determine the authenticity of a questioned sample.

\subsection{Forensic Loss in the Feature Embedding Space}
\label{subsec:ForensicLoss}

General Euclidean distance has been widely adopted in comparing similarity between a pair of samples in the embedding space \cite{roth2020revisiting}.
However, Euclidean distance merely captures the equally weighted dimension-by-dimension distances between the two sets of features.
The calculation of Euclidean distance assumes that different dimensions in a feature vector are uncorrelated and of the same variance \cite{bateni2020improved}.
Such distance metric may not be suitable to a document recapture detection task with large sample variation.
For example, an unseen recaptured sample may have a smaller Euclidean distance with a genuine sample in the training set than that with a recaptured one.
The design of a suitable distance metric in the embedding space is important.

Mayer and Stamm designed a CNN-based subnet in measuring the ``forensic'' similarity or distance between two samples in the embedding space \cite{mayer2020forensic}.
This network is with a simple two-layer structure, including operations that map the input features to a high dimensional space, such as element-wise multiplication and concatenation.
A similarity score is output at the final layer.
A high similarity score indicates that the two input samples have a similar forensic trace, while a low similarity score implies that different forensic traces.
However, the score is thresholded to 0 and 1 in computing the cross-entropy loss.
Such binary labelling strategy is prone to over-fitting to the training samples (as demonstrated in our experiment).

To mitigate such issue, we propose to take advantages of both approaches.
The forensic similarity function is incorporated into a set of general metric learning loss.
Thus, the distances between two samples are measured by a trainable subnet, and the similarity score will be considered as a continuous distance metric.

In the following, we adopt a set of general metric learning loss in \cite{perez2019deep} with the forensic similarity function \cite{mayer2020forensic}, yielding the proposed triplet similarity loss \cite{hu2014discriminative}.
\begin{align}
\label{eqn:TripletLoss}
\mathcal{L}_{ts} =& \sum_{i=1}^b \left[ e^{1-S[f(x_{r_i}), f(x_{p_i})]} - e^{1-S[f(x_{r_i}), f(x_{n_i})]} + \gamma \right]_+ \nonumber \\
                 =& e\cdot \sum_{i=1}^b \left[ e^{-S[f(x_{r_i}), f(x_{p_i})]} - e^{-S[f(x_{r_i}), f(x_{n_i})]} + \frac{\gamma}{e} \right]_+ \nonumber \\
                 \stackrel{(a)}{=}& \sum_{i=1}^b \left[ e^{-S[f(x_{r_i}), f(x_{p_i})]} - e^{-S[f(x_{r_i}), f(x_{n_i})]} + \gamma' \right]_+
\end{align}
where $b$ is the size of a mini-batch, $\gamma$ is the a hyper-parameter that defines the safety margin between two sets of distance, $[ \cdot ]_+$ is the hinge function that accepts an operand only if it is non-negative, and {\small$S(\cdot , \cdot)$} evaluates the similarity between two samples with the deep network adopted from \cite{mayer2020forensic}.
There are a few points to be noted in our triplet similarity loss defined in Eq.~(\ref{eqn:TripletLoss}).
First, the similarity score {\small$S(\cdot , \cdot)$} ranges from 0 to 1 indicates low and high similarity, respectively.
Therefore, {\small$1-S(\cdot , \cdot)$} measures the disparity between a sample pair.
Such disparity is then scaled by an exponential function which nonlinearly up-weights the samples with large distances.
Second, Eq.~(\ref{eqn:TripletLoss}-a) is achieved by eliminating a constant scaling factor $e$ from the loss function and by setting {\small$\gamma'=\gamma/e$}.
After simplification, we see that the proposed triplet similarity loss encourages a large value of {\small$e^{-S[f(x_{r_i}), f(x_{n_i})]}$}, i.e., a low similarity score between {\small$f(x_{r_i})$} and {\small$f(x_{n_i})$}, while punishes a large value of {\small$e^{-S[f(x_{r_i}), f(x_{p_i})]}$}.
Thus, the positive samples will be clustered around the references, and the negative samples will be pushed away.

Besides, we also consider the relative distance between positive and negative samples with the normalized softmax loss \cite{movshovitz2017no}.
\begin{align}
\label{eqn:SoftmaxLoss}
\mathcal{L}_{ns} =& -\sum_{i=1}^b \log \left[ \frac{e^{S[f(x_{r_i}), f(x_{p_i})]-1}}{e^{S[f(x_{r_i}), f(x_{p_i})]-1}+e^{S[f(x_{r_i}), f(x_{n_i})]-1}} \right] \nonumber \\
                 \stackrel{(a)}{=}& -\sum_{i=1}^b \log \left[ \frac{e^{S[f(x_{r_i}), f(x_{p_i})]}}{e^{S[f(x_{r_i}), f(x_{p_i})]}+e^{S[f(x_{r_i}), f(x_{n_i})]}} \right] \nonumber \\
                 \stackrel{(b)}{=}& \sum_{i=1}^b \log \left[ 1 + \frac{e^{S[f(x_{r_i}), f(x_{n_i})]}}{e^{S[f(x_{r_i}), f(x_{p_i})]}} \right]
\end{align}
where Eq.~(\ref{eqn:SoftmaxLoss}-a) is achieved by eliminating a constant scaling factor $e^{-1}$ from both the nominator and denominator.
The proposed normalized softmax loss is in a similar form with the proxy-NCA loss \cite{movshovitz2017no}. Eq.~(\ref{eqn:SoftmaxLoss}-b) rewrites the loss in a straightforward form.
It indicates that, for each triplet, the minimum loss is $\log (1+e^{-1})$.
The normalized softmax loss encourages a small {\small$S[f(x_{r_i}), f(x_{n_i})]$} and a large {\small$S[f(x_{r_i}), f(x_{p_i})]$} in a non-linear scale.

Overall, the proposed forensics loss can be written as
\begin{align}
\label{eqn:ForensicLoss}
\mathcal{L}_{fl} = \mathcal{L}_{ts} + \alpha \cdot \mathcal{L}_{ns}
\end{align}
where $\alpha$ is a hyper-parameter that specifies the weights of different loss components.

\subsection{Triple Sampling Strategy}
\label{subsec:TripleSampling}

Employing all possible triplets to train our model is neither efficient nor effective \cite{roth2020revisiting}.
As demonstrated in Section~\ref{subsubsec:ExperimentInterset}, randomly sampling triplets from all samples introduces about 3.5\% performance degradation compared to the proposed sampling strategy.
Triple sampling strategy focus on mining representative triplets from all possible combination.
Inspired by the workflow of a document forensic expert, who determines the authenticity of a signature by comparing the questioned specimen with a genuine one \cite{allen2015foundations}, we select a high-quality genuine document image as the reference sample in a triplet.
To allow a more efficient training process, we generate triples from all samples by minimizing the variations, such as differences in document contents, acquisition conditions, and resolutions.

First, the triples are only generated among the document images from the same template (i.e., the same type of student ID card in our database) such that the image patches in each triple are with similar contents.
It should be noted that the images in a triplet should be normalized to the same resolution to allow generation of patches with similar content, such as a face region as shown in Fig~\ref{fig:NetworkArchitecture}.

Second, only the samples with similar resolutions are chosen to form a triple. This is to avoid the quality loss in some heavily downsampled images.
For example, if a scanned high resolution document image is compared with a captured low resolution one, heavy downsampling operation will be performed on the high resolution one to allow patch-to-patch comparison at the same location of two samples.
However, such downsampling operation introduces unnecessary difficulty in extracting useful forensic traces.
Thus, the low- and high-resolution triplets are generated from samples collected by mobile phones and scanners, respectively, to avoid heavy downsampling.

Third, we also follow \cite{ferreira2017data, mayer2020forensic} in selecting discriminative image patches to train our model. As shown in the triplet selection stage of Fig.~\ref{fig:NetworkArchitecture}, the discarded patches (highlighted in yellow) are those regions dominated by uniform backgrounds, and the patches with texts and graphics are more informative.

Finally, the semi-hard triplet mining strategy \cite{schroff2015facenet} is also adopted in our triple sampling process to allow a more efficient training process.

\begin{figure}
\centerline{\includegraphics[width=3.5in]{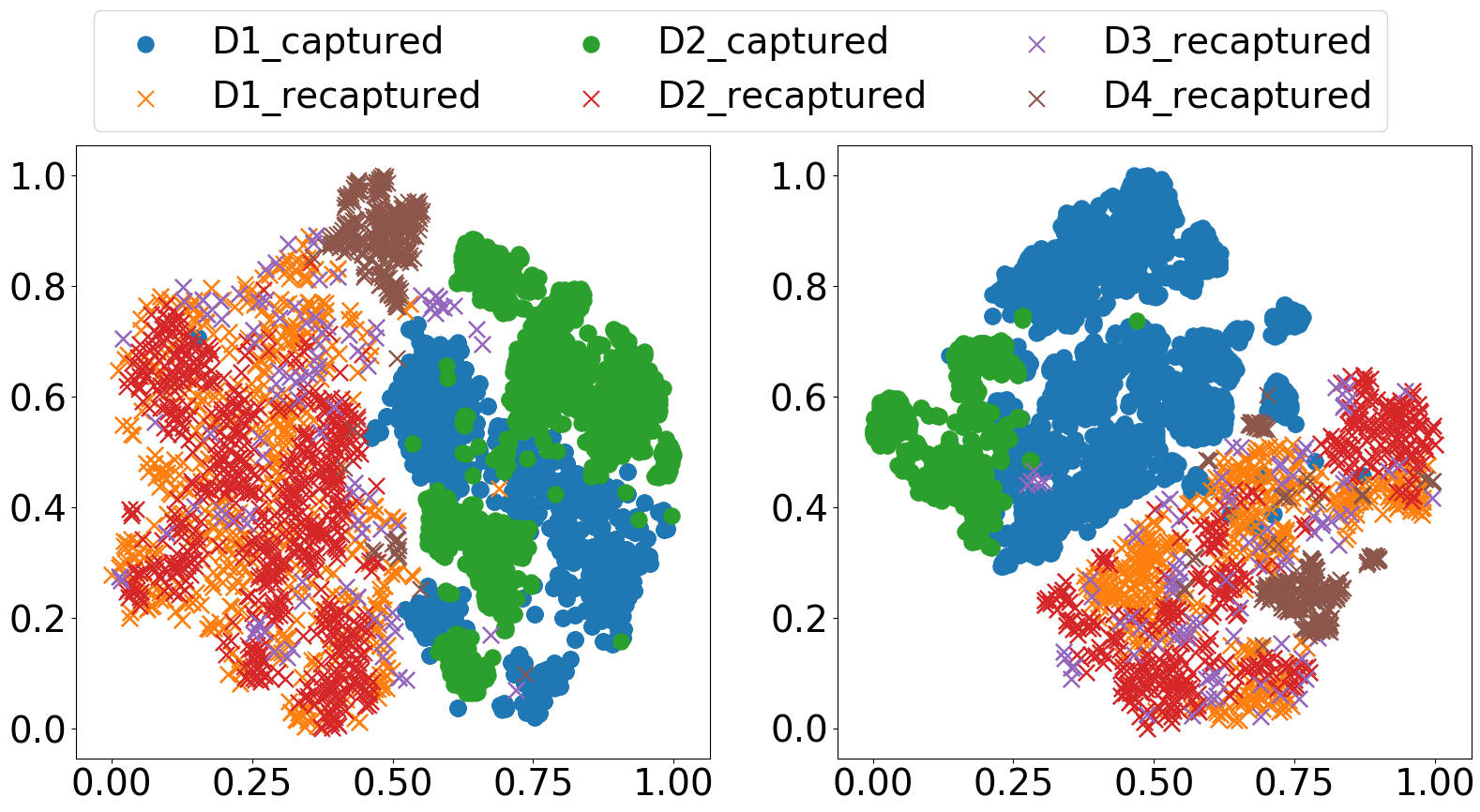}}
\centerline{\footnotesize(a) \hspace{1.6in} (b)}
\vspace{-0.2cm}
\caption{The t-SNE visualizations of the extracted features by the proposed scheme (with ResNet50 backbone). (a) using dataset $D_1$ as training set, and $D_2, D_3, D_4$ as testing sets. (b) using dataset $D_2$ as training set, and $D_1, D_3, D_4$ as testing sets. $D_1$ and $D_2$ are two datasets that collected by two sets of different devices (printers, cameras and scanners). $D_3$ contains recaptured document images produced with high-quality glossy photo paper. $D_4$ denotes the dataset obtained from display-and-capture channel.}
\label{fig:TSNE}
\vspace{-0.25cm}
\end{figure}

\subsection{Visualization of Feature Embedding Space}
\label{subsec:tSNE}

As shown in Fig.~\ref{fig:TSNE}, the resulting feature embedding separates the captured and recaptured samples in two different classes.
To plot the t-SNE figures \cite{van2008visualizing}, we employ two different datasets captured and recaptured by different printing and imaging devices to train our deep model.
These two training datasets are denoted as $D_1$ and $D_2$ according to the discussion in Section~\ref{subsec:Database}.
The visualization shows how samples are separated in a two-dimensional space.

There are some interesting points to be noted.
First, the captured and recaptured samples (blue and red solid circles) in $D_1$ and $D_2$ are clustered together though they are collected by two sets of different devices.
It demonstrates that good generalization performance of the proposed deep model towards different devices, regardless of the training datasets.
Second, most of the recaptured samples from an LCD screen (denoted as $D_4$ as discussed in Section~\ref{subsubsec:ExperimentITW}) are found to be in a single cluster marked by brown crosses.
It indicates that the proposed scheme not only identifies the recaptured samples from an LCD screen, but also separates samples from different recapturing channels (i.e., print-and-capture and display-and-capture).
Third, there are some purple crosses (high-quality recaptured samples with glossy photo paper, denoted as $D_3$ in Section~\ref{subsubsec:ExperimentITW}) among the blue and red solid circles (captured images in $D_1$ and $D_2$).
Such phenomena does imply the difficulties in authenticating the high-quality recaptured samples.

\section{Database and Experimental Results}
\label{sec:DataResult}

\subsection{Our Captured and Recaptured Document Image Database}
\label{subsec:Database}

\begin{table*} [t]
\begin{center}
\caption{The devices used for collecting dataset $D_1$ and $D_2$, which contains 672 samples and 432 samples, respectively.}
\footnotesize
\label{tab:Dataset1and2}
\begin{tabular}{| c | c | c |  c | c | c |} \hline
Set & \multicolumn{2}{c|}{1st Imaging Device} & Printer & \multicolumn{2}{c|}{2nd Imaging Device} \\ \hline \hline
\multirow{7}{*}{$D_1$} & \multirow{4}{*}{Phone} & XiaoMi 8 (12 MP) & & \multirow{4}{*}{Phone} & XiaoMi 8 (12 MP) \\ \cline{3-3} \cline{6-6}
  & & RedMi Note 5 (12 MP) &  & & RedMi Note 5 (12 MP) \\ \cline{3-3} \cline{6-6}
  & & Huawei P9 (12 MP) &  & & Huawei P9 (12 MP) \\ \cline{3-3} \cline{6-6}
  & & Apple iPhone 6 (8 MP) & HP OfficeJet 258 & & Apple iPhone 6 (8 MP) \\ \cline{2-3} \cline{5-6}
  & \multirow{3}{*}{Scanner} & Brother DCP-1519 (1200 DPI) & ($4800 \times 1200$ DPI) & \multirow{3}{*}{Scanner} & Brother DCP-1519 (1200 DPI) \\ \cline{3-3} \cline{6-6}
  & & Epson V330 (1200 DPI) &  & & Epson V330 (1200 DPI) \\ \cline{3-3} \cline{6-6}
  & & Benq K810 (1200 DPI) & & & Benq K810 (1200 DPI) \\ \hline \hline
\multirow{4}{*}{$D_2$} & \multirow{2}{*}{Phone} & Apple iPhone 6s (12 MP) & HP LaserJet m176n & \multirow{2}{*}{Phone} & Apple iPhone 6s (12 MP) \\ \cline{3-3} \cline{6-6}
  & & Oppo Reno (48 MP) & ($2400 \times 2400$ DPI) & & Oppo Reno (48 MP) \\ \cline{2-6}
 & \multirow{2}{*}{Scanner} & Epson V850 (3200 DPI) & Epson L805 & \multirow{2}{*}{Scanner} & Epson V850 (3200 DPI) \\ \cline{3-3} \cline{6-6}
  & & HP LaserJet m176n (1200 DPI) & ($5760 \times 1440$ DPI) & & HP LaserJet m176n (1200 DPI) \\ \hline
\end{tabular}
\end{center}
\vspace{-0.25cm}
\end{table*}

\begin{figure}[t]
\centerline{\includegraphics[width=3.25in]{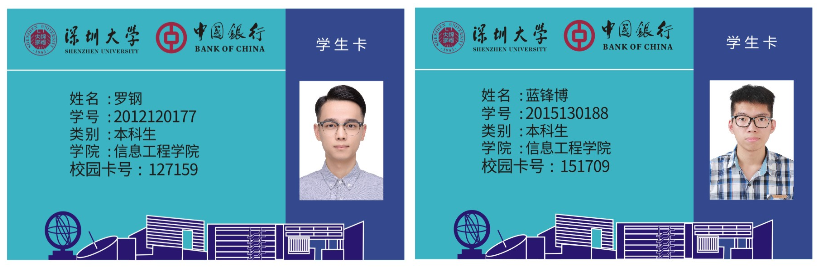}}
\vspace{-0.1cm}
\centerline{\footnotesize(a)}
\vspace{0.1cm}
\hspace{-0.05cm}
\centering{
\begin{minipage}[c]{.45\linewidth}
  \centering
  \centerline{\includegraphics[width=1.62in]{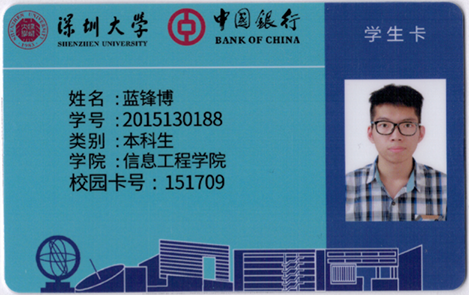}}
  \centerline{\footnotesize(b)}
\end{minipage}
\begin{minipage}[c]{.45\linewidth}
  \centering
  \centerline{\includegraphics[width=1.62in]{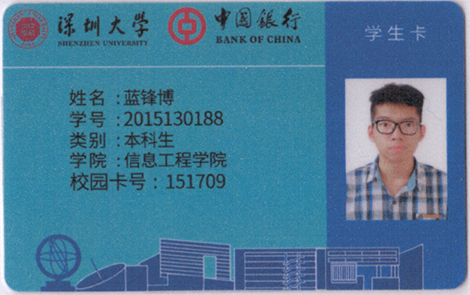}}
  \centerline{\footnotesize(c)}
\end{minipage}
}
\vspace{-0.1cm}
\caption{(a) The original ID images synthesized with Adobe CorelDRAW for our experiment. (b) Genuine document image collected by Brother DCP-1519 scanner. (c) Recaptured document collected by Brother DCP-1519 scanner (both first and second imaging process) and HP OfficeJet 258 printer.}
\label{fig:OriginalID}
\vspace{-0.25cm}
\end{figure}

To investigate the problem of document recapturing detection, a database consists of captured and recaptured document images is needed.
We collect and share publicly a high-quality dataset for this problem.
First and foremost, the content of our documents should be chosen carefully.
Some legal documents (e.g., passport, ID card, certificate), which contain sensitive privacy, are not suitable to be shared in the public domain.
Without loss of generality, student ID cards from 5 universities are chosen as the original document images in our experiment.
As shown in Fig.~\ref{fig:OriginalID}-(a), these images are synthesized with Adobe CorelDRAW software according to the templates available in the public domain.
The genuine documents are then manufactured by a specialized third-party on acrylic plastic to preserve the details in the graphical contents.

\begin{figure}[t!]
\centerline{\includegraphics[width=3.4in]{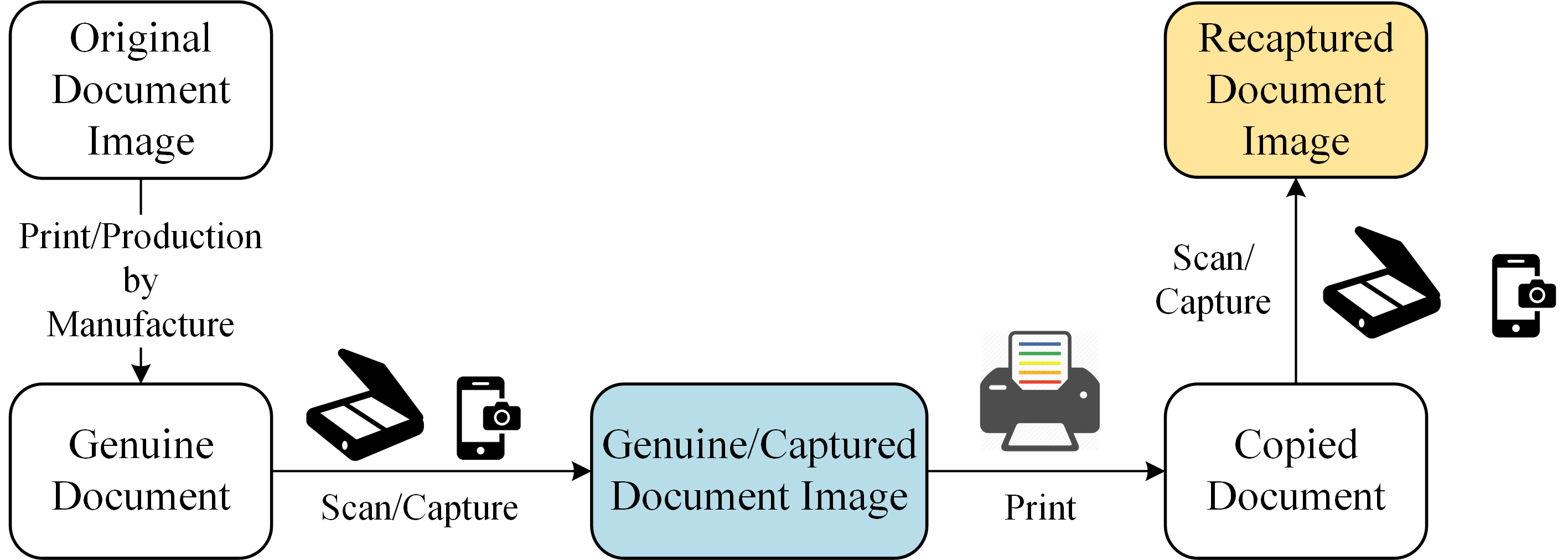}}
\caption{The block diagram of collecting genuine/captured document images and recaptured document images.}
\label{fig:ImageRecapturing}
\vspace{-0.25cm}
\end{figure}

Our database contains two datasets.
Dataset I ($D_1$) collects 84 genuine and 588 recaptured document images which are captured or recaptured, respectively, by $7 \times 7$ different combinations of devices.
As shown in Fig.~\ref{fig:ImageRecapturing}, the original document is printed by an authorized party to generate the genuine document, which are then scanned/captured to yield the captured document images.
To collect the recaptured document images, the copied document images are print and re-acquired (by scanner or camera).
Dataset II ($D_2$) follows the same data collection procedure but employs a different set of devices.
It contains 48 genuine and 384 recaptured document images.
As shown in Table~\ref{tab:Dataset1and2}, we have employed 4 phones, 3 scanners and 1 printer in collecting $D_1$, while 2 phones (including a high-quality camera phone, Oppo Reno with resolution of 48 mega-pixels (MP)), 2 scanners (including a high-end scanner, Epson V850 with optical resolution of 6400 dots per inch (DPI)), and 2 printers (including a high-end printer, Epson L805 with resolution of 5760$\times$1440 DPI) in gathering $D_2$.
Thus, dataset II considers the recapturing attack with high-quality off-the-shelf devices.

To collect a high-quality image set, there are a few thumb of rules in the data collection process of $D_1$ and $D_2$.
\begin{itemize}
\item \emph{Camera Phones}: point perpendicular towards the document plane to avoid geometric distortion, set to the highest supported resolution, and the captured images are saved in JPEG format with the highest quality factor;
\item \emph{Environmental Light}: evenly illuminated by a lamp to avoid shadowing;
\item \emph{Scanners}: set to a resolution of 1200 DPI, except the Epson V850 remains at its default value of 3200 dpi, and the scanned output is saved in TIF format;
\item \emph{Printers}: set to color mode with the finest printing resolution;
\item \emph{Printing Substrate}: office paper with 120 g/m$^2$.
\end{itemize}

Two typical example of genuine and recaptured student ID images are shown in Fig.~\ref{fig:OriginalID}-(b) and (c), respectively.
It can be observed by comparing both images that the recapturing process introduces noise, blurring and color distortion.

\subsection{Experimental Results}
\label{subsec:Result}

Based on the dataset presented in Sec.~\ref{subsec:Database}, we evaluate the performance of the proposed method and the benchmarking methods, including traditional handcrafted feature-based classifier (Local Binary Pattern \cite{ojala2002multiresolution} + Support Vector Machine, LBP+SVM), generic CNN models (e.g., ResNet \cite{he2016deep}, ResNeXt \cite{xie2017aggregated}, DenseNet \cite{huang2017densely}), and a state-of-the-art document recapturing detection scheme (the recurrent Comparator with attention models (CRC+Attn) \cite{berenguel2019recurrent}).
Ablation studies on the forensic similarity (FS) subnet and triplet loss (TL) are also performed to demonstrate the advantages over \cite{mayer2020forensic} (a Siamese network with FS subnet for general image forensic) and \cite{perez2019deep} (a metric learning approach for face spoofing detection with similar triplet loss).
To quantitatively compare different schemes, the performance metrics Attack Presentation Classification Error Rate (APCER) and Bona Fide Presentation Classification Error Rate (BPCER), Equal Error Rate (EER), and Area Under the ROC Curve (AUC) are employed in our experiments.

The implementation details of different approaches are provided as follow.

\emph{LBP+SVM}: We employ the LBP implementation in Matlab (version 2018b) with the default parameters to extra the $\mbox{LBP}_8^1$ feature vector. The resulting feature vectors of image samples are then fed into the support vector machine (SVM) \cite{chang2011libsvm} with linear and radial bias function (RBF) kernels.

\emph{Generic CNN models}: We select 3 representative generic CNN models with different layer configurations, i.e., ResNet with 34, 50, 101 and 152 layers, ResNeXt with 50 and 101 layers, as well as DenseNet with 121, 169 and 201 layers.
These CNN models are pre-trained with the ImageNet dataset and are fine-tuned with our document recapturing dataset to achieve a better performance \cite{yosinski2014transferable}.
The original fully-connected (FC) layers of the pre-trained networks are replaced by two FC layers (with 256 nodes in the last layer) since our task only involves samples from 2 classes, instead of 1000 classes in the original ImageNet dataset.
During our fine-tuning process, we freeze the weights of the shallow part of the network (layers before the FC layers) and uses our database data to retrain the weights of the FC layers.
The input sample resolution of these networks is $224\times224$ pixels.
The batch size is 128, and the model is trained with the Adam optimizer.
The training period is 20 epochs with a learning rate of $1 \times 10^{-4}$ and the cross entropy loss.
Our implementation is based on Pytorch 1.10 and Tensorflow 1.13.1.

\emph{CRC+Attn}: We adopt the implementation provided in \cite{berenguel2019recurrent}.
The recurrent comparator with attention model (CRC+Attn) is considered in our comparison since another configuration (i.e., recurrent comparator with co-attention, CRC+Co-Attn) suffers from the same limitation.
Different from the data pre-processing in \cite{berenguel2019recurrent}, pixel-precise alignment cannot be guaranteed in our dataset due to two reasons.
First, the paired samples are not necessarily from the same ID image with the same content.
More commonly, they are generated from ID images from different subjects.
As shown in Fig.~\ref{fig:NetworkArchitecture}, some key user information and personal photo are different, and pixel-precise alignment is not possible.
Second, some samples are acquired by mobile cameras which introduce lens, geometric, and surface distortions.
It is not easy to align the details of two documents with pixel precision.

\begin{table} [t!]
\begin{center}
\caption{Experimental result in Dataset $D_1$ and $D_2$. The samples in each dataset are divided in an 8:1:1 ratio for the training, validation and testing sets, respectively. The best performance under each condition is bold-faced.}
\footnotesize
\label{tab:Dataset1Result}
\begin{tabular}{| c | c | c | c | c |} \hline
\multirow{2}{*}{Methods} &  \multicolumn{2}{c|}{$D_1$} & \multicolumn{2}{c|}{$D_2$} \\ \cline{2-5}
                 & EER      & AUC    & EER    & AUC \\ \hline \hline
LBP+SVM (Linear) &  23.08\%	& 0.8204 & 22.24\% & 0.8380  \\ \hline
LBP+SVM (RBF)	& 25.00\%	& 0.8145 & 27.72\% & 0.8056  \\ \hline
DenseNet121		& \textbf{0.01\%}	& \textbf{1.0000}   & \textbf{0.01\%} & \textbf{1.0000}  \\ \hline
DenseNet169		& \textbf{0.01\%}	& \textbf{1.0000}   & \textbf{0.01\%} & \textbf{1.0000}  \\ \hline
DenseNet201		& \textbf{0.01\%}	& \textbf{1.0000}   & \textbf{0.01\%} & \textbf{1.0000}  \\ \hline
ResNeXt50		& \textbf{0.01\%}	& \textbf{1.0000}   & \textbf{0.01\%} & \textbf{1.0000}  \\ \hline
ResNeXt101		& \textbf{0.01\%}	& \textbf{1.0000}   & \textbf{0.01\%} & \textbf{1.0000}  \\ \hline
ResNet34		& \textbf{0.01\%}	& \textbf{1.0000}   & \textbf{0.01\%} & \textbf{1.0000}  \\ \hline
ResNet50		& \textbf{0.01\%}	& \textbf{1.0000}   & \textbf{0.01\%} & \textbf{1.0000}  \\ \hline
ResNet101		& \textbf{0.01\%}	& \textbf{1.0000}   & \textbf{0.01\%} & \textbf{1.0000}  \\ \hline
ResNet152		& \textbf{0.01\%}	& \textbf{1.0000}   & \textbf{0.01\%} & \textbf{1.0000}  \\ \hline
CRC+Attn \cite{berenguel2019recurrent} & 24.47\% & 0.8193 & 18.84\% & 0.8938 \\ \hline
DenseNet121+FS  & \textbf{0.01\%} & \textbf{1.0000} & \textbf{0.01\%} & \textbf{1.0000} \\ \hline
ResNeXt101+FS   & \textbf{0.01\%} & \textbf{1.0000} & \textbf{0.01\%} & \textbf{1.0000} \\ \hline
ResNet50+FS     & \textbf{0.01\%} & \textbf{1.0000} & \textbf{0.01\%} & \textbf{1.0000} \\ \hline
DenseNet121+TL  & \textbf{0.01\%} & \textbf{1.0000} & \textbf{0.01\%} & \textbf{1.0000}  \\ \hline
ResNeXt101+TL    & \textbf{0.01\%} & \textbf{1.0000} & \textbf{0.01\%} & \textbf{1.0000} \\ \hline
ResNet50+TL      & \textbf{0.01\%} & \textbf{1.0000} & \textbf{0.01\%} & \textbf{1.0000} \\ \hline
DenseNet121+Proposed    & \textbf{0.01\%} & \textbf{1.0000} & \textbf{0.01\%} & \textbf{1.0000} \\ \hline
ResNeXt101+Proposed     & \textbf{0.01\%} & \textbf{1.0000} & \textbf{0.01\%} & \textbf{1.0000} \\ \hline
ResNet50+Proposed       & \textbf{0.01\%} & \textbf{1.0000} & \textbf{0.01\%} & \textbf{1.0000} \\ \hline
\end{tabular}
\end{center}

\vspace{0.15cm}
\footnotesize{- \textbf{CNN+FS}: adopts Siamese network and the Forensic Similarity subnet \cite{mayer2020forensic}; \\
- \textbf{CNN+TL}: employs Siamese network and the Triplet Loss in \cite{perez2019deep};\\
- \textbf{CNN+Proposed}: the proposed approach that combines both FS and TL.}
\vspace{-0.25cm}
\end{table}

\emph{CNN+FS}: We employ the forensic similarity (FS) subnet \cite{mayer2020forensic} to evaluate the distances between the features extracted by the CNN models.
However, there are several distinct differences between the architecture proposed in \cite{mayer2020forensic} and our implementation.
First, the fine-tuned generic CNN backbones (ResNet, ResNeXt and DenseNet) are adopted in our implementation to extract the recapturing traces, such as halftone texture and color degradation, while the MISLnet CNN \cite{bayar2018constrained} is employed in \cite{mayer2020forensic} to extract image manipulation traces in digital domain.
The network architecture in our implementation allows a fair and straightforward way of identifying the advantage of the FS subnet.
Second, instead of using a fix decision threshold on the similarity score (0.5, as shown in \cite{mayer2020forensic}), we determine the threshold by picking a value that optimizes the training accuracy.
Third, the triplet sampling strategy in our implementation is different.
The original implementation randomly choosing image patches with half of patch pairs chosen from the same and different camera models, respectively, while we generate triplets following the description in Sec.~\ref{subsec:TripleSampling} by considering the uniqueness of our problem.

\emph{CNN+TL}: We use the triplet loss (TL) proposed in \cite{perez2019deep} to train the Siamese-based CNNs.
However, the original and our implementation of TL are different mainly in the two aspects.
First, the triplets in \cite{perez2019deep} are generated through the semi-hard negative mining, while our implementation further considers the contents and resolutions of different document images as elaborated in Sec.~\ref{subsec:TripleSampling}.
Second, for fair comparison, we only utilize samples from the source domain to determine the authenticity of a questioned sample in the testing process, instead of computing a posteriori probability with samples in the target domain except for the results in Table~\ref{tab:CrossFixThresholdContentFT}.

\emph{CNN+Proposed}: We take advantages of both FS and TL as proposed in Sec.~\ref{sec:Proposed}. This approach follows similar training procedure of the generic CNN models. The hyper-parameters are set as $\gamma=0.2$ and $\alpha=0.3$, which are the same as the settings in CNN+TL approach. Moreover, to demonstrate the effectiveness of the proposed triplet sampling strategy proposed in Section~\ref{subsec:TripleSampling}, we performance an ablation study by employing the random sampling strategy. The proposed network with random sampling strategy is denoted as CNN+Proposed (R).

To investigate the challenges posed by recapturing attacks, we compare the performances of the proposed scheme with some benchmarking methods on our database under different experimental protocols.
\begin{itemize}
\item \emph{Intra-dataset Experiment}: The training and testing images are from the same dataset ($D_1$ or $D_2$) acquired by the same set of devices;
\item \emph{Cross-dataset Experiment}: The training and testing images are from different datasets which involves different printing and imaging devices;
\item \emph{In-the-wild Experiment}: The training and testing images are from different printing substrates, different recapturing channels, and different types of documents.
\end{itemize}

\subsubsection{Intra-dataset Experiment}
\label{subsubsec:ExperimentIntraset}

\begin{table} [t]
\begin{center}
\caption{Cross dataset evaluation in $D_1$ and $D_2$. $D_1 \rightarrow D_2$ denotes training and testing with dataset $D_1$ and $D_2$, respectively, while $D_2 \rightarrow D_1$ means otherwise. The best performance for each backbone network is bold-faced.}
\label{tab:CrossDataset}
\footnotesize
\begin{tabular}{| c | c | c | c | c |} \hline
\multirow{2}{*}{Methods} & \multicolumn{2}{c|}{$D_1 \rightarrow D_2$} & \multicolumn{2}{c|}{$D_2 \rightarrow D_1$} \\ \cline{2-5}
            & EER       & AUC       & EER       & AUC \\ \hline \hline
DenseNet121	& 14.40\% 	& 0.9293 	& 8.32\% 	& 0.9731  \\ \hline
DenseNet169	& 17.39\%	& 0.9130 	& 7.14\%	& 0.9822  \\ \hline
DenseNet201	& 19.64\%	& 0.8960 	& 13.46\% 	& 0.9363  \\ \hline
ResNeXt50	& 21.73\%	& 0.8597 	& 11.90\% 	& 0.9487  \\ \hline
ResNeXt101	& 10.85\% 	& 0.9654 	& 7.14\% 	& 0.9795  \\ \hline
ResNet34	& 17.28\%	& 0.8978 	& 14.28\% 	& 0.9333  \\ \hline
ResNet50	& 13.04\%	& 0.9440 	& 16.66\% 	& 0.9432  \\ \hline
ResNet101	& 24.34\% 	& 0.8268 	& 9.20\% 	& 0.9756  \\ \hline
ResNet152	& 24.87\%	& 0.7987 	& 4.75\% 	& 0.9908  \\ \hline
DenseNet121+FS  & \textbf{2.61\%} & \textbf{0.9976} & 3.75\% & 0.9955 \\ \hline
ResNeXt101+FS   & 4.95\% & 0.9942 & 3.74\% & 0.9940 \\ \hline
ResNet50+FS     & 2.34\% & 0.9976 & \textbf{3.40\%} & \textbf{0.9961} \\ \hline
DenseNet121+TL   & 14.31\% & 0.9351 & 20.80\% & 0.9024 \\ \hline
ResNeXt101+TL    & 14.28\% & 0.9367 & 17.14\% & 0.9065 \\ \hline
ResNet50+TL      & 12.52\% & 0.9395 & 12.51\% & 0.9483 \\ \hline
DenseNet121+Proposed    & {3.40\%} & {0.9936} & \textbf{1.50\%} & \textbf{0.9985} \\ \hline
ResNeXt101+Proposed     & \textbf{2.27\%} & \textbf{0.9988} & \textbf{1.47\%} & \textbf{0.9978} \\ \hline
ResNet50+Proposed       & \textbf{1.30\%} & \textbf{0.9997} & 4.00\% & {0.9935} \\ \hline
DenseNet121+Proposed (R)    & {6.51\%} & {0.9865} & {4.28\%} & {0.9930} \\ \hline
ResNeXt101+Proposed (R)     & {6.86\%} & {0.9904} & {7.15\%} & {0.9814} \\ \hline
ResNet50+Proposed (R)       & {6.79\%} & {0.9891} & 5.67\% & {0.9879} \\ \hline
\end{tabular}
\end{center}
\vspace{0.15cm}
\footnotesize{- \textbf{CNN+Proposed (R)}: the proposed approach with random sampling strategy.}
\vspace{-0.35cm}
\end{table}

\begin{figure}[t!]
\centerline{
\includegraphics[width=3.5in]{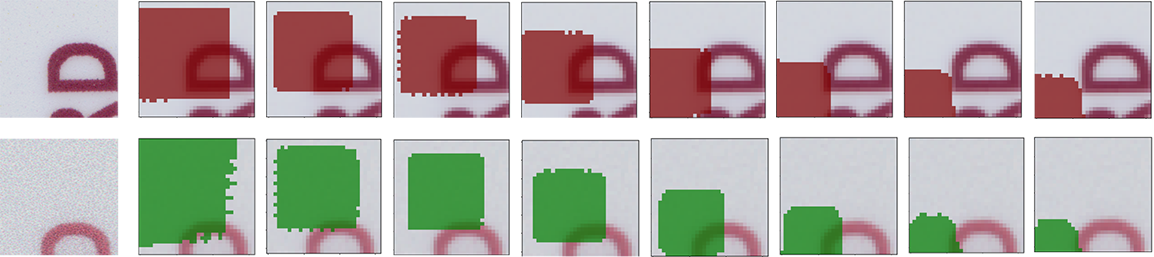}}
\centerline{\footnotesize (a) samples with authentication error.}
\vspace{0.25cm}
\centerline{
\includegraphics[width=3.5in]{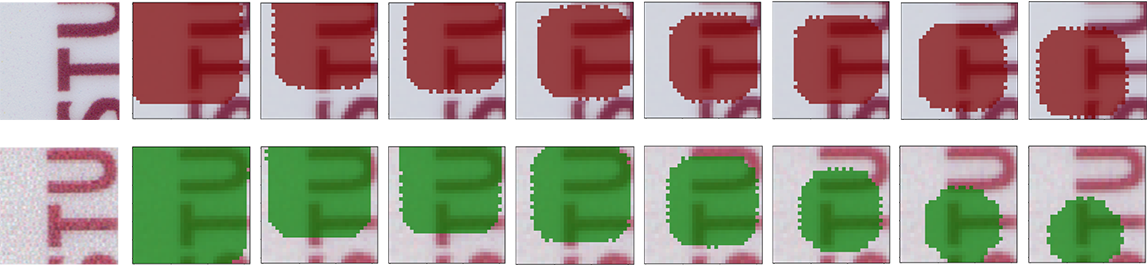}}
\centerline{\footnotesize (b) samples with correct authentication result.}
\caption{The iterative process proposed in \cite{berenguel2019recurrent} that searchs for attention regions in a pair of patches of size $224 \times 224$ pixels. (a) The recurrent comparator \cite{berenguel2019recurrent} fails to locate the corresponding attention regions in two patches due to the poor alignment. (b) The recurrent comparator \cite{berenguel2019recurrent} locates the corresponding attention regions in two patches with similar contents.}
\label{fig:CRCAttn}
\vspace{-0.35cm}
\end{figure}

In this experiment, the samples in each dataset are divided in an 8:1:1 ratio for training, validation, and testing sets, respectively.
As shown in Table~\ref{tab:Dataset1Result}, performances of the generic recapture detection framework are satisfactory in the scenarios where training and testing data is sampled from the same subset.
For example, most of the CNN-based schemes achieve EER = 0.01\% and AUC = 1.0000 in the experiment conducted within $D_1$ and $D_2$, respectively.
This experiment shows baseline performance of the CNN-based schemes under an ideal scenario.

However, there are several approaches which perform less satisfactorily.
On the one hand, the LBP-based classifiers with default parameters achieve AUC less than 0.85.
This is due to the limited descriptive power of the LBP features and the shallow structure of a SVM classifier.
On the other hand, the AUCs of CRC+Attn are 0.81 and 0.89 in the Intra-dataset experiment of $D_1$ and $D_2$, respectively, which is a significant performance degradation compared to the original performance reported in \cite{berenguel2019recurrent}.
Through investigation, we have found that this is because of the difference between the dataset in \cite{berenguel2019recurrent} and our datasets in Sec.~\ref{subsec:Database}.
The triplet selection procedure in \cite{berenguel2019recurrent} selects the patches at the exact same position of the document, which requires precise alignment between different image samples.
However, such precise alignment is not guaranteed in our datasets to reflect a more practical setting.
As shown in Fig.~\ref{fig:CRCAttn}, the mis-alignment among the input patches has introduced difficulties in identifying attention regions with the recurrent comparator which inspects the patches iteratively.
It should be noted that such limitation is also observed in the CRC+Co-Attn approach \cite{berenguel2019recurrent}.

\subsubsection{Cross-dataset Experiment}
\label{subsubsec:ExperimentInterset}

\begin{table*} [t]
\begin{center}
\caption{Cross dataset evaluation in $D_1$ and $D_2$ with decision thresholds of different schemes set by BPCER = 1\%, 5\%, and 10\%, respectively. APCER is employed as the performance metric. The best performance for each backbone network is bold-faced.}
\label{tab:CrossFixThreshold}
\footnotesize
\begin{tabular}{| c | c | c | c | c | c | c |} \hline
\multirow{2}{*}{Methods} & \multicolumn{3}{c|}{$D_1 \rightarrow D_2$} & \multicolumn{3}{c|}{$D_2 \rightarrow D_1$} \\ \cline{2-7}
                 & BPCER=1\% & BPCER=5\% & BPCER=10\% & BPCER=1\% & BPCER=5\% & BPCER=10\% \\ \hline \hline
DenseNet121      & 28.65\%  & 24.48\%   & 14.32\%   & 19.39\%   & 14.28\%   & 7.14\%    \\ \hline
ResNeXt101   	 & 20.05\%  & 16.67\%  & 9.64\%   & 11.90\%   & 11.05\%   & 4.76\%    \\ \hline
ResNet50     	 & 33.59\%   & 20.83\%   & 13.02\%   & 29.93\%   & 26.53\%   & 24.83\%    \\ \hline
DenseNet121+FS   & \textbf{2.34\%}  & \textbf{0.00\%}   & \textbf{0.00\%}   & 4.76\%   & 2.04\%   & 0.85\%    \\ \hline
ResNeXt101+FS    & 3.65\%  & 0.78\%   & {0.00\%}  & 2.55\%   & 1.36\%   & 1.02\%    \\ \hline
ResNet50+FS      & 4.43\%  & 1.30\%   & {0.00\%}   & \textbf{3.91\%}   & \textbf{1.02\%}   & \textbf{0.85\%}    \\ \hline
DenseNet121+TL   & 20.57\%   & 20.57\%   & 7.55\%   & 21.43\%   & 20.41\%   & 16.33\%    \\ \hline
ResNeXt101+TL    & 10.93\%   & 10.68\%   & 9.90\%   & 23.12\%   & 16.67\%   & 12.76\%    \\ \hline
ResNet50+TL      & 12.50\%   & 12.50\%   & 11.20\%   & 22.96\%  & 18.71\%   & 16.16\%    \\ \hline
DenseNet121+Proposed    & {4.95\%}   & {3.39\%}   & 2.60\%   & \textbf{0.85\%}   & \textbf{0.17\%}   & \textbf{0.17\%}   \\ \hline
ResNeXt101+Proposed     & \textbf{1.30\%}  & \textbf{0.52\%}   & \textbf{0.00\%}   & \textbf{1.19\%}   & \textbf{0.85\%}   & \textbf{0.34\%}    \\ \hline
ResNet50+Proposed       & \textbf{1.30\%}  & \textbf{0.00\%}   & \textbf{0.00\%}   & {4.59\%}   & {3.57\%}   & {2.04\%}    \\ \hline
\end{tabular}
\end{center}
\vspace{-0.35cm}
\end{table*}

\begin{table*} [t]
\begin{center}
\caption{Evaluation on high-quality printing substrate of the deep models trained by $D_1$ and $D_2$, respectively. The decision thresholds are set by BPCER = 1\%, 5\%, and 10\% during the training of models presented in Table~\ref{tab:CrossFixThreshold}. APCER is employed as the performance metric. The best performance for each backbone network is bold-faced.}
\label{tab:CrossFixThresholdSubstrate}
\footnotesize
\begin{tabular}{| c | c | c | c | c | c | c |} \hline
\multirow{2}{*}{Methods} & \multicolumn{3}{c|}{Same Devices ($D_2 \rightarrow D_3$)} & \multicolumn{3}{c|}{Different Devices ($D_1 \rightarrow D_3$)} \\ \cline{2-7}
                 & BPCER=1\% & BPCER=5\% & BPCER=10\% & BPCER=1\% & BPCER=5\% & BPCER=10\% \\ \hline \hline
DenseNet121		 & 38.75\%  & 36.25\%   & 23.75\%   & 40.00\%   & 33.75\%   & 20.00\%    \\ \hline
ResNeXt101		 & 23.75\%  & 23.75\%   & 11.25\%   & 37.50\%   & 33.75\%   & 25.00\%    \\ \hline
ResNet50		 & 50.00\%  & 50.00\%   & 50.00\%   & 45.00\%   & 40.00\%   & 33.75\%    \\ \hline
DenseNet121+FS   & 33.75\%  & 25.00\%   & 23.50\%   & 18.75\%   & 18.75\%   & 10.00\%    \\ \hline
ResNeXt101+FS    & 32.50\%  & 21.50\%   & 18.75\%   & 33.75\%   & 28.75\%   & 27.50\%    \\ \hline
ResNet50+FS      & 35.00\%  & 25.00\%   & 10.00\%   & 13.75\%   & 10.00\%   & 10.00\%    \\ \hline
DenseNet121+TL   & 25.00\%  & 21.25\%   & 17.50\%   & 43.75\%   & 32.50\%   & 17.50\%    \\ \hline
ResNeXt101+TL    & 23.75\%  & 20.00\%   & 12.50\%   & 21.25\%   & 18.75\%   & 17.50\%    \\ \hline
ResNet50+TL      & 27.50\%  & 21.25\%   & 11.25\%   & 26.25\%   & 23.75\%   & 18.75\%    \\ \hline
DenseNet121+Proposed    & \textbf{6.25\%}  & \textbf{5.00\%}   & \textbf{1.25\%}   & \textbf{12.50\%}   & \textbf{8.75\%}   & \textbf{5.00\%}    \\ \hline
ResNeXt101+Proposed     & \textbf{7.50\%}  & \textbf{3.75\%}   & \textbf{1.25\%}   & \textbf{7.50\%}   & \textbf{6.25\%}  & \textbf{2.50\%}    \\ \hline
ResNet50+Proposed       & \textbf{6.25\%}  & \textbf{2.50\%}   & \textbf{2.50\%}   & \textbf{5.00\%}   & \textbf{3.75\%}   & \textbf{3.75\%}    \\ \hline
\end{tabular}
\end{center}
\vspace{-0.35cm}
\end{table*}

We also evaluate the cross-dataset performances of different approaches (excluding LBP and CRC+Attn with poor performances in the intra-dataset experiment).
The two datasets $D_1$ and $D_2$ are collected with two sets of different devices while the image content and acquisition environment (such as illumination conditions) remain the same.
It can be observed in Table~\ref{tab:CrossDataset} that the recapture detection performances of the generic CNN-based schemes (ResNet, ResNeXt and DenseNet) degrade significantly when the training and testing data is inhomogeneous.
The EERs of most generic CNN-based schemes are more than 10\% (except DenseNet121, DenseNet169, ResNeXt101 and ResNet101 under $D_2 \rightarrow D_1$), which are over 10\% increment compared to those of the cross-dataset experiment.
Moreover, it is interesting to see that the performances under $D_2 \rightarrow D_1$ scenarios are better than those of $D_1 \rightarrow D_2$, except ResNet50.
This is due to two reasons.
First, the images dataset $D_2$ are collected with both laserjet and inkjet printers, while dataset $D_1$ only involves an inkjet printer.
Second, the devices used in $D_2$ are of higher quality, such as Epson L805 with $5760 \times 1440$ DPI, Oppo Reno with 48 MP, and Epson V850 with 3200 DPI.
Comparing the two experiments ($D_1 \rightarrow D_2$ and $D_2 \rightarrow D_1$), it is easier to learn discriminative features from $D_2$ and to obtain better performances on a lower quality dataset $D_1$.

\begin{figure}[t!]
\centerline{\includegraphics[width=3.5in]{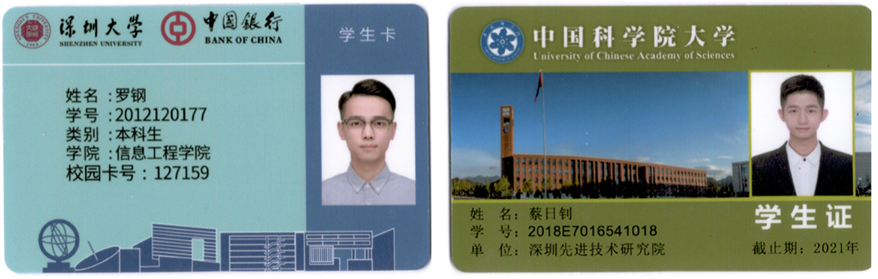}}
\caption{The high-quality recaptured document images with glossy photo paper by HP LaserJet M176n scanner (both first and second imaging process) and Epson L805 printer.}
\vspace{-0.25cm}
\label{fig:PrintingSubstrate}
\end{figure}

\begin{figure}[t!]
\centerline{\includegraphics[width=3.5in]{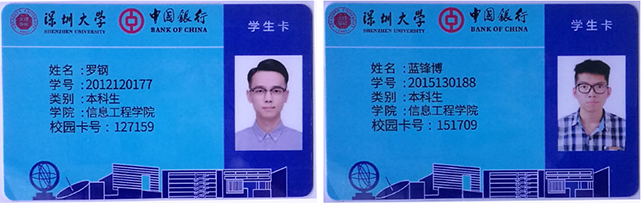}}
\caption{The recaptured document images from display-and-capture channel by Brother DCP-1519 scanner (the first imaging process), the LCD screen of ThinkPad X1, and Huawei P9 phone camera (the second imaging process).}
\vspace{-0.25cm}
\label{fig:DCrecaptured}
\end{figure}

For the sake of comparison, only the CNN architecture with the best performance, i.e., ResNet50, RseNeXt101 and DenseNet121 are selected for our following comparison.
Generally speaking, the CNN+FS and the proposed schemes work much better than the other approaches.
EERs of the CNN+FS and the proposed schemes are less than 5\%.
Meanwhile, the CNN+TL schemes do not have significant improvement over the generic CNN-based schemes.
This can be explained by the Euclidean distance employed in \cite{perez2019deep} which is not a suitable metric for measuring the discrepancy between different samples in an cross-dataset setting.
On contrary, the FS network \cite{mayer2020forensic} measures the sample distance with an end-to-end trainable network and leads to good performances for CNN+FS and the proposed schemes.

Moreover, to show the system performance under a practical setting, we evaluate the APCERs of different schemes while keeps the BPCERs at 1\%, 5\% and 10\%, respectively.
As shown in Table~\ref{tab:CrossFixThreshold}, only the CNN+FS and the proposed schemes keep APCERs within 10\% under pre-determined security levels set by different BPCERs.
Similar to the results in Table~\ref{tab:CrossDataset}, the proposed approach outperforms CNN+FS under most conditions, except for DenseNet121 backbone under $D_1 \rightarrow D_2$ and ResNet50 backbone under $D_2 \rightarrow D_1$.

Last but not least, we performance an ablation study on the proposed triplet sampling strategy as described in Section~\ref{subsec:TripleSampling}.
The random sampling strategy, denoted by `R' in the last three rows of Table~\ref{tab:CrossDataset}, is evaluated with the same number of triplets as the CNN+proposed configuration.
On average, over 3.50\% increment is observed in the EERs compared to the implementation with the proposed triplet sampling strategy.

\subsubsection{In-the-wild Experiment}
\label{subsubsec:ExperimentITW}

In this subsection, we investigate the performances of different schemes under some more practical scenarios where the training and testing images are from different printing substrates, different recapturing channels, and different types of documents.

\begin{table*} [t]
\begin{center}
\caption{Evaluation on different recapturing channels of the deep models trained by $D_1$ and $D_2$, respectively. The decision thresholds are set by BPCER = 1\%, 5\%, and 10\% during the training of models presented in Table~\ref{tab:CrossFixThreshold}. APCER is employed as the performance metric. The best performance for each backbone network is bold-faced.}
\label{tab:CrossFixThresholdDC}
\footnotesize
\begin{tabular}{| c | c | c | c | c | c | c |} \hline
\multirow{2}{*}{Methods} & \multicolumn{3}{c|}{Same Imaging Devices ($D_1 \rightarrow D_4$)} & \multicolumn{3}{c|}{Different Imaging Devices ($D_2 \rightarrow D_4$)} \\ \cline{2-7}
                   & BPCER=1\% & BPCER=5\% & BPCER=10\% & BPCER=1\% & BPCER=5\% & BPCER=10\% \\ \hline \hline
DenseNet121        & 100.00\%   & 100.00\%    & 100.00\%   & 95.38\%   & 94.05\%   & 82.74\%    \\ \hline
ResNeXt101         & 100.00\%   & 100.00\%    & 99.40\%   & 98.17\%   & 98.17\%   & 86.13\%     \\ \hline
ResNet50           & 100.00\%   & 100.00\%    & 99.40\%   & 100.00\%    & 100.00\%    & 100.00\%    \\ \hline
DenseNet121+FS     & 21.43\%  & 12.50\%   & 9.52\%    & 36.90\%   & 18.45\%   & 13.69\%    \\ \hline
ResNeXt101+FS      & 23.21\%  & 16.07\%   & 9.52\%    & 25.00\%   & 15.48\%   & 0.59\%    \\ \hline
ResNet50+FS        & 28.57\%  & 17.26\%   & 10.12\%   & 30.36\%   & 16.67\%   & 12.50\%    \\ \hline
DenseNet121+TL     & 15.48\%  & 11.90\%   & 8.93\%    & 25.60\%   & 16.67\%   & 11.31\%    \\ \hline
ResNeXt101+TL      & 21.43\%  & 16.07\%   & 12.50\%   & 19.64\%   & 13.69\%   & 0.833\%    \\ \hline
ResNet50+TL        & 11.31\%  & 10.71\%   & 6.55\%    & 19.05\%   & 12.50\%   & 10.12\%    \\ \hline
DenseNet121+Proposed    & \textbf{12.50\%}  & \textbf{7.74\%}   & \textbf{5.36\%}   & \textbf{0.60\%}   & \textbf{0.60\%}   & \textbf{0.00\%}    \\ \hline
ResNeXt101+Proposed     & \textbf{3.57\%}   & \textbf{2.38\%}   & \textbf{0.60\%}   & \textbf{1.19\%}   & \textbf{1.19\%}   & \textbf{0.00\%}   \\ \hline
ResNet50+Proposed       & \textbf{8.93\%}   & \textbf{5.95\%}   & \textbf{5.36\%}   & \textbf{2.98\%}   & \textbf{2.38\%}   & \textbf{2.38\%}    \\ \hline
\end{tabular}
\end{center}
\vspace{-0.35cm}
\end{table*}

\vspace{0.25cm}
\noindent $\bullet$ \emph{Different Printing Substrates}

Different printing substrates renders different printing distortions due to the different ink-substrate interaction characteristics \cite{yang2003ink}.
As shown in Fig.~\ref{fig:PrintingSubstrate}, the ink spreading can be better controlled by employing a high-quality printing substrate.
Compared to the recaptured sample in Fig.~\ref{fig:OriginalID}-(c), these high-quality samples in Fig.~\ref{fig:PrintingSubstrate} contain less noise.
To evaluate the robustness of the proposed scheme under different printing substrates, we employ iPhone 6s, HP LaserJet M176n and Epson L805 (belong to the set of devices in $D_2$) and glossy photo paper to collect 80 high-quality recaptured samples.
This dataset is denoted as $D_3$.

In this part, the pre-trained models obtained from Table~\ref{tab:CrossFixThreshold} are tested against the recaptured samples in $D_3$.
It should be noted that all the samples in $D_3$ are recaptured samples, and the performance metric for this case is APCER.
According to different training sets ($D_1$ or $D_2$) used, this experiment includes scenarios with same or different recapturing devices, i.e., $D_2 \rightarrow D_3$ or $D_1 \rightarrow D_3$.
In general, the scenarios with different devices are more challenging than those with the same devices.

It can be seen in Table~\ref{tab:CrossFixThresholdSubstrate} that the proposed schemes outperform CNN+FS and CNN+TL under all settings.
Let us focus on the performance under BPCER=5\% with the same devices, the APCERs of proposed schemes are less than 5\% which shows at least 75\% improvement over those of CNN+TL and CNN+TS.

\vspace{0.25cm}
\noindent $\bullet$ \emph{Different Recapturing Channels}

As is well-known that the recapturing attack can also be carried out with the display-and-capture channel \cite{cao2010identification}.
Such recapturing attack with an LCD screen is very different from the reprinting process considered in our datasets $D_1$ and $D_2$.
Comparing Fig.~\ref{fig:OriginalID}-(c) and Fig.~\ref{fig:DCrecaptured}, we see that there is less distortion in noise (mainly introduced by printing process) but heavier distortion in color in the recaptured samples from an LCD screen.
In this part, we consider a cross domain testing.
The training samples are collected by printers and scanners, while those in the testing set are recaptured from an LCD screen.
To facilitate our study, we collect a high-quality dataset following the setting in \cite{thongkamwitoon2015image} to avoid visible moir\'e pattern.
In this part, we use two smart phones (Huawei P9 and iPhone 6) in dataset $D_1$ since this dataset contains more genuine document images (to be displayed in the recapturing process) than $D_2$.
Totally, 168 recaptured images from an LCD screen from ThinkPad X1 with $1920 \times 1080$ pixels resolution is gathered.
This dataset is denoted as $D_4$.
According to different training set ($D_1$ or $D_2$) used, this experiment includes scenarios with same or different recapturing devices, i.e., $D_1 \rightarrow D_4$ or $D_2 \rightarrow D_4$.

It can be seen in Table~\ref{tab:CrossFixThresholdDC} that the generic CNN models (DenseNet121, ResNeXt101 and ResNet50) are not able to generalize towards different recapturing channels.
This is because the recapturing artifacts in printing and displaying devices are different, but the generic CNN models have only been trained to classify the artifacts generated in the printing process, such as halftone patterns.
For the samples recaptured from displaying devices, no halftone pattern can be observed due to the limited display resolution.
Therefore, over 94\% of the recaptured samples are mistaken as genuine samples under 0.05 BPCER by the generic CNN models.

\begin{figure}[t]
\centering{
\begin{minipage}[c]{.45\linewidth}
  \centering
  \centerline{\includegraphics[width=1.62in]{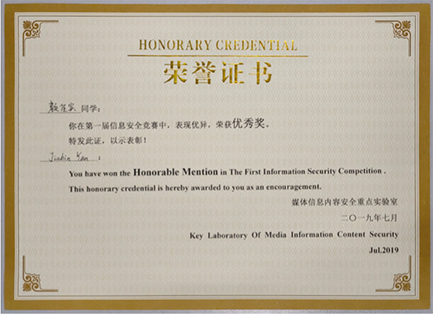}}
  \centerline{\footnotesize(a)}
\end{minipage}
\hfil
\begin{minipage}[c]{.45\linewidth}
  \centering
  \centerline{\includegraphics[width=1.62in]{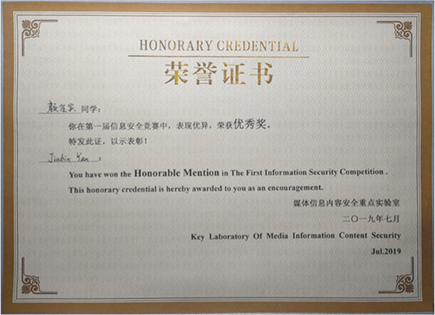}}
  \centerline{\footnotesize(b)}
\end{minipage}
}
\vspace{-0.1cm}
\caption{Samples of certificate document images. (a) a genuine certificate image captured by Oppo Reno. (b) a recaptured certificate image collected by Oppo Reno (both first and second imaging process) and Epson L805.}
\label{fig:CertificateImage}
\vspace{-0.25cm}
\end{figure}

The proposed schemes work well under this cross domain testing with APCERs lower than 0.10 except the case with DenseNet121 under 0.01 BPCER and the same imaging devices $D_1 \rightarrow D_4$.
It is not straightforward that the proposed schemes perform much better under the scenarios with different imaging devices than those with the same imaging devices.
By analyzing the results in both classes, we found that the error samples are mainly high-quality recapturing images which are similar to the samples reprinted by a high-quality printer (Epson L805) in $D_2$.
Thus, pre-training the model with $D_2$ leads to a higher performance among the high-quality recapturing samples.

\begin{table*} [t]
\begin{center}
\footnotesize
\caption{Evaluation on different document content of the deep models trained by $D_1$ and $D_2$, respectively. The decision thresholds are set by BPCER = 1\%, 5\%, and 10\% during the training process.}
\vspace{-0.4cm}
\label{tab:CrossFixThresholdContent}
\resizebox{\textwidth}{!}{%
\begin{tabular}{| c | c | c | c | c | c | c | c | c | c | c | c | c |} \hline
\multirow{3}{*}{Methods} & \multicolumn{6}{c|}{Same Devices ($D_2 \rightarrow D_5$)} & \multicolumn{6}{c|}{Different Devices ($D_1 \rightarrow D_5$)} \\ \cline{2-13}
  & \multicolumn{2}{c|}{BPCER=1\%} & \multicolumn{2}{c|}{BPCER=5\%} & \multicolumn{2}{c|}{BPCER=10\%} & \multicolumn{2}{c|}{BPCER=1\%} & \multicolumn{2}{c|}{BPCER=5\%} & \multicolumn{2}{c|}{BPCER=10\%} \\ \cline{2-13}
  & APCER & BPCER & APCER & BPCER & APCER & BPCER & APCER & BPCER & APCER & BPCER & APCER & BPCER   \\ \hline \hline
DenseNet121+Proposed    & 15.00\%  & 23.33\%   & 10.00\%  & 26.67\%   & 5.00\%   &30.00\%   & 3.33\%    & 73.33\%   & 3.33\%    & 73.33\%   & 3.33\%   & 80.00\%     \\ \hline
ResNeXt101+Proposed     & 0.00\%   & 50.00\%   & 0.00\%   & 50.00\%   & 0.00\%   &50.00\%   & 20.00\%   & 30.00\%   & 15.00\%   & 43.33\%   & 1.67\%   & 53.33\%     \\ \hline
ResNet50+Proposed       & 8.33\%   & 50.00\%   & 6.67\%   & 50.00\%   & 6.67\%   &50.00\%   & 10.00\%   & 63.33\%   & 6.00\%    & 70.00\%   & 0.00\%   & 70.00\%     \\ \hline
\end{tabular}
}
\end{center}
\vspace{-0.35cm}
\end{table*}

\begin{table*} [t]
\begin{center}
\footnotesize
\caption{Evaluation on different document types of the deep models trained by $D_1$ and $D_2$, respectively, and fine-tuned by 6 triplets in the target domain. The decision thresholds are set by BPCER = 1\%, 5\%, and 10\% during the training process.}
\vspace{-0.4cm}
\label{tab:CrossFixThresholdContentFT}
\resizebox{\textwidth}{!}{%
\begin{tabular}{| c | c | c | c | c | c | c | c | c | c | c | c | c |} \hline
\multirow{3}{*}{Methods} & \multicolumn{6}{c|}{Same Devices ($D_2 \rightarrow D_5$)} & \multicolumn{6}{c|}{Different Devices ($D_1 \rightarrow D_5$)} \\ \cline{2-13}
& \multicolumn{2}{c|}{BPCER=1\%} & \multicolumn{2}{c|}{BPCER=5\%} & \multicolumn{2}{c|}{BPCER=10\%} & \multicolumn{2}{c|}{BPCER=1\%} & \multicolumn{2}{c|}{BPCER=5\%} & \multicolumn{2}{c|}{BPCER=10\%} \\ \cline{2-13}
                        & APCER & BPCER & APCER & BPCER & APCER & BPCER & APCER & BPCER & APCER & BPCER & APCER & BPCER   \\ \hline \hline
DenseNet121+Proposed    & 10.00\%    & 2.78\%    & 6.67\%    & 8.33\%    & 6.67\%    & 11.11\%   & 11.67\%    & 5.56\%    & 6.67\%    & 8.33\%     & 3.22\%    & 16.67\%  \\ \hline
ResNeXt101+Proposed     & 3.33\%     & 5.56\%    & 3.33\%    & 11.11\%   & 0.00\%    & 11.11\%   & 8.33\%     & 11.11\%   & 6.67\%    & 11.11\%    & 5.00\%    & 13.89\%  \\ \hline
ResNet50+Proposed       & 13.33\%    & 5.56\%    & 5.00\%    & 5.56\%    & 3.33\%    & 8.33\%    & 11.67\%    & 5.56\%    & 8.33\%    & 11.11\%    & 6.67\%    & 11.11\%  \\ \hline
\end{tabular}
}
\end{center}
\vspace{-0.35cm}
\end{table*}

\vspace{0.25cm}
\noindent $\bullet$ \emph{Different Types of Documents}

A practical document authentication system processes different types of document images, such as ID card, certificate and other legal documents.
It is necessary to evaluate a cross domain scenarios where the training and testing samples from different types of documents.
Worth noting that, this scenario not only considers different document contents, but also include the variations in different printing substrates (e.g., the acrylic plastic of ID cards and office paper of other documents), as well as different printing techniques.
To facilitate this study, we have collected some document images with the same printing (Epson L805 and HP LaserJet m176n) and imaging devices (HP LaserJet m176n, Epson V850, and Oppo Reno) as $D_2$.
Samples of certificate document images are shown in Fig.~\ref{fig:CertificateImage}.
This dataset, denoted as $D_5$, contains 36 and 60 genuine and recaptured certificate images in 6 different templates.
According to different training set ($D_1$ or $D_2$) used, this experiment includes scenarios with same or different devices, i.e., $D_2 \rightarrow D_5$ or $D_1 \rightarrow D_5$.

This experiment is conducted exclusively on the proposed scheme since the generic CNNs, CNN+FS and CNN+TL have failed in the less challenging tests on printing substrates and recapturing channels.
During the testing process, the template of a questioned document is new to the training data, three pairs of positive and negative samples in the target domain are employed.
This can be considered as a few shot learning setting where only a few samples in the target domain are available.

As shown in Table~\ref{tab:CrossFixThresholdContent}, evaluations of the pre-trained models show that the BPCERs of different backbone networks are higher than 0.5 in most case.
It reflects a high authentication error rate for the genuine document images.
This is due to the triplet loss and forensic similarity network are not generalizable across different types of documents.
More specifically, the genuine documents of ID cards and certificates are manufactured by different printing techniques and different printing substrates.
The two types of genuine documents are therefore with distinctive features.

To further evaluate this scenario, we select 6 triplets (18 images) in two different certificate templates to fine-tune our models.
To balance the samples (i.e., the number of patches) from different types of devices, 2 and 4 triplets from scanners and mobile cameras are selected.
After fine-tuning with 6 triplets (18.75\% of the samples in $D_5$), we observe significant performance improvement in BPCERs.
Specifically, the BPCER of the proposed scheme with ResNet50 backbone (at 5\% BPCER) has decreased from 50\% and 70\% to 11.11\% and 11.11\%, respectively, for the experimental scenarios with the same and different devices.

\section{Conclusion}
\label{sec:Conclusion}

In this work, we have proposed a recaptured document detection scheme by taking advantages of both metric learning and image forensics techniques.
Experimental results have demonstrated that the proposed scheme has good generalization performance under variations of printing (inkjet and laser printers)/imaging (high-quality scanners and low-quality camera phones) devices, substrates (general office paper and glossy photo paper) and recapturing channels (print-and-scan and display-and-capture).
The proposed scheme is simple but competitive.
Moreover, we have presented the first publicly available image dataset for document recapturing problem covering practical but challenging situations.

In the future, the remaining issues in the scenario of different document types (i.e., the third bullet in Sec.~\ref{subsubsec:ExperimentITW}) should be further investigated in two aspects.
First, the samples in target domain used to fine-tune this model should be reduced or be eliminated to improve the practicality of our approach.
Second, it is also beneficial to study a scenario where the reference image in the target domain is not available.

\ifCLASSOPTIONcaptionsoff
  \newpage
\fi

\bibliographystyle{IEEEtran}
\bibliography{DocumentRecapture}

\end{document}